\documentclass[twocolumn,showpacs,preprintnumbers,amsmath,amssymb,superscriptaddress]{revtex4}

\usepackage{amsfonts}
\usepackage{mathrsfs}
\usepackage{amsmath}
\usepackage{amssymb}
\usepackage{txfonts}
\usepackage{graphicx}
\usepackage{bm}
\usepackage{color}
\usepackage{subfigure}
\begin{document}



\title{A simplified method for calculating the ac Stark shift of hyperfine levels}

\author{Xia Xu}
\affiliation{School of Electronics Engineering $\&$ Computer
Science, Peking University, Beijing 100871, China}
\author{Bo Qing}
\affiliation{Department of Physics, Tsinghua University, Beijing
100084, China}
\author{Xuzong Chen}
\affiliation{School of Electronics Engineering $\&$ Computer
Science, Peking University, Beijing 100871, China}
\author{Xiaoji Zhou}\thanks{Electronic address: xjzhou@pku.edu.cn }
\affiliation{School of Electronics Engineering $\&$ Computer
Science, Peking University, Beijing 100871, China}




\begin{abstract}
The ac Stark shift of hyperfine levels of neutral atoms can be
calculated using the third order perturbation theory(TOPT), where
the third order corrections are quadratic in the atom-photon
interaction and linear in the hyperfine interaction. In this paper,
we use Green's function to derive the $E^{[2+\epsilon]}$ method
which can give close values to those of TOPT for the differential
light shift between two hyperfine levels. It comes with a simple
form and easy incorporation of theoretical and experimental atomic
structure data. Furthermore, we analyze the order of approximation
and give the condition under which $E^{[2+\epsilon]}$ method is
valid.
\end{abstract}


\maketitle

\section{Introduction}

The recent developments in precision measurement~\cite{Godone,Simon}
and optical communication~\cite{Souza} require a possible way to
calculate the ac Stark shift with considerable precision. In many
cases, the second order perturbation theory(SOPT)~\cite{Grimm},
which is capable of utilizing the existing theoretical and
experimental atomic structure data, is used to compute the light
shifts. For instance, in today's researches on atomic clocks, it has
been realized that the accuracy and stability can be substantially
improved by trapping cold atoms in a standing wave of light (optical
lattice)~\cite{Katori,Ye,Blatt,Ludlow}. Because of the minimization
of Doppler and recoil effects, light shift caused by trapping laser
is essential. Therefore the wavelength of the trapping laser should
be tuned to a region where the light shifts of the two clock
transition states cancel each other out. This wavelength is called
``magic wavelength''~\cite{Archive}. In optical clocks and terahertz
clocks, the clock transition is between the fine structure of atomic
ground state and excited states, and we can utilize the SOPT to
compute the light shift of the clock transition. The shift arises in
the second order of perturbation theory which is quadratic in the
electric field strength. The calculations suggest the existence of
magic wavelength both for optical-clock
transitions~\cite{Wilpers,Barbaer,Barber,Gao} and terahertz-clock
transitions~\cite{terahertz}.

Recently, the alkali-metal atom like rubidium (Rb) and cesium (Cs)
are considered as potential choices for microwave lattice clocks,
using the two field insensitive hyperfine levels of the ground state
as clock transition levels~\cite{rev}. However, because SOPT doesn't
take into account the hyperfine interaction, the results are
identical for the hyperfine doublet of the ground state at arbitrary
values of trapping laser wavelength. To solve this problem, the
third order perturbation theory (TOPT)~\cite{P.Rosenbusch} was
proposed by extending the formalism to the higher order of
perturbation theory, and the third order corrections are quadratic
in the field amplitude and linear in the hyperfine interaction. The
theory requires using \textit{ab initio} approach to construct the
atomic structure database. Here we introduce the $E^{[2+\epsilon]}$
method which takes the hyperfine interactions into consideration in
SOPT. We will show that for a wide range of trapping laser
wavelength, $E^{[2+\epsilon]}$ method gives close results to those
of TOPT and experiments. In addition, $E^{[2+\epsilon]}$ method
comes with a simple form, easy incorporation of theoretical and
experimental atomic structure data, and therefore is more applicable
for other group elements.

The remainder of this manuscript is organized as follows. We use
Green's function and diagrammatic representation to derive the
$E^{[2+\epsilon]}$ method in Sec.II. In Sec.III, the differential
light shifts between two field insensitive hyperfine levels of the
ground state of Cs and Rb are calculated using both methods, due to
their potential application in microwave lattice clocks. In the
calculation, besides utilizing the existing experimental atomic
structure data, we use GraspVU code~\cite{Parpia} to construct our
own database of atomic structure, which is summarized in Appendix A.
The discussions and conclusions are given in Sec.IV.

\section{$E^{[2+\epsilon]}$ method}


\subsection{Hyperfine structure}
We start with no light fields. The Hamiltonian $\hat{h}$ of the
system can be written as the sum of the unperturbed part $\hat{h_0}$
and the perturbation $\Delta \hat{h}$.
\begin{eqnarray}
\begin{split}
\hat{h}&=\hat{h_0} + \Delta \hat{h},\\
\hat{h_0}&=\hat{H}_{fs},\\
\Delta \hat{h}&=\hat{H}_{hfs} .\\
\end{split}
\label{hamiltonian1}
\end{eqnarray}
Here $\hat{h_0}$ is the fine structure Hamiltonian $\hat{H}_{fs}$,
and the perturbation $\Delta \hat{h}$ is the hyperfine interaction
Hamiltonian $\hat{H}_{hfs}$. In the coupled representation, the
eigenstate of $\hat{H}_{fs}$ can be written as
\begin{eqnarray}
|n I J  FM_{F}\rangle=\sum\limits_{\scriptstyle{M_J, M_I}} {\langle
J M_J I M_I | F M_F\rangle \times |nJM_{J}\rangle|IM_{I}\rangle},
\label{hyperfinestate}
\end{eqnarray}
where $n$ is the principle quantum number, $I$ is the nuclear spin,
$J$ is the electronic total angular momentum and $F=I+J$ is the
total angular momentum; $M_I$, $M_J$ and $M_F$ are the projections
of $I$, $J$ and $F$ on the quantization axis, respectively. $\langle
J M_J I M_I | F M_F\rangle$ is the Clebsh-Gordan coefficient.
However, Eq. (\ref{hyperfinestate}) is not an eigenstate of
$\hat{h}$, because the hyperfine interactions have non-zero
off-diagonal matrix elements. In the following, we use a shorthand
notation $|i\rangle\equiv|n_i I J_i F_i M_{Fi}\rangle$ for
convenience.

\begin{figure}[h]
\includegraphics [height=3cm,width=8cm]{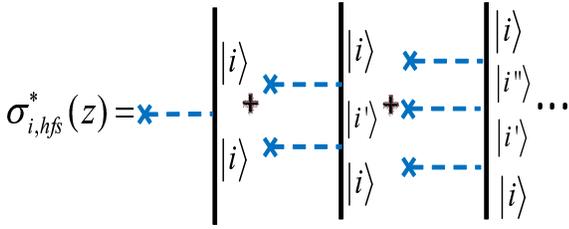}
\caption{ (Color online)Diagrammatic representation of the proper
self-energy $\sigma^*_{i,hfs}(z)$. $|i'\rangle$ and $|i''\rangle$
denote different eigenstates of $\hat{H_{fs}}$ which have the same
parity with $|i\rangle$ but with different $J$ or principle quantum
numbers. Every solid line(black) marked by $|i\rangle$ represents a
factor $g^{0}_{i}(z)=1/(z-\langle i|\hat{H_{fs}}|i\rangle)$, Every
dashed line(blue) marked by $|i\rangle$ and $|i'\rangle$ denotes a
factor $\langle i|\hat{H}_{hfs}|i'\rangle$. A summation is performed
over the index $i',i''...$ in the end.} \label{rep_diagram1}
\end{figure}

The Green's function of $\hat{h}$ with complex variable $z$ is
\begin{eqnarray}
\begin{split}
g_{n_i,i}(z)&=\langle i|\frac{1}{z-\hat{h}}|i\rangle\\
&=\frac{1}{z-{E_{i,fs}}-\langle i|\sigma^*_{i,hfs}(z)|i\rangle},
\end{split}
\label{greenfunction1}
\end{eqnarray}
where $\sigma^*_{i,hfs}(z)$ is the proper self-energy. which can be
diagrammatically represented by the infinite sum in Fig.
(\ref{rep_diagram1}). Because the hyperfine state energy
$E_{i}=E_{i,fs}+E_{i,hfs}$ is one pole of the Green's function,
where $E_{i,fs}$ and $E_{i,hfs}$ are the fine structure energy and
the hyperfine corrections, respectively, we have:
\begin{eqnarray}
\begin{split}
E_{i,hfs}=\sigma^*_{i,hfs}(E_{i,fs}+E_{i,hfs})=\sigma^*_{i,hfs}(E_{i}).
\end{split}
\label{thezrelation}
\end{eqnarray}\\


\subsection{The ac Stark shift}
Now we consider a neutral atom in a far-off-resonance laser field
with frequency $\nu = \omega/2\pi$. The laser field is assumed to be
in a Fock state $|R\rangle=|N\rangle (N\gg1)$, where $N$ equals the
mean photon number. The Hamiltonian $\hat{H}$ of the system can be
written as the sum of the unperturbed part $\hat{H_0}$ and the
perturbation $\Delta \hat{H}$:
\begin{eqnarray}
\begin{split}
\hat{H}&=\hat{H_0} + \Delta \hat{H},\\
\hat{H_0}&=\hat{H}_{R}+\hat{H}_{fs},\\
\Delta \hat{H}&=\hat{H}_{hfs} +\hat{H}_e.\\
\end{split}
\label{hamiltonian}
\end{eqnarray}
Here $\hat{H_0}$ consists of the radiation field Hamiltonian
$\hat{H}_{R}$ and the fine structure Hamiltonian $\hat{H}_{fs}$. The
state $|i;N\rangle\equiv|i\rangle\otimes|N\rangle$ is an eigenstate
of $\hat{H}_0$ with eigenvalue ${E_{i,fs}}+N\hbar\omega$.


The perturbation $\Delta \hat{H}$ of Eq.(\ref{hamiltonian}) takes
into account the hyperfine interaction $\hat{H}_{hfs}$ and
atom-photon interaction $\hat{H}_e$. For $\hat{H}_e$, we use the
dipole approximation $\hat{H}_e = - \hat{d} \cdot
\vec{\varepsilon}$, where $\hat{d}$ is the electric dipole moment
and $\vec{\varepsilon}$ is the electric field vector. Here for the
sake of simplicity, we have ignored the atom's external degree of
freedom.

The Green's function of $\hat{H}$ with complex variable $z$ has a
similar form to Eq. (\ref{greenfunction1}):
\begin{eqnarray}
\begin{split}
G_{i;N}(z)&=\langle i;N|\frac{1}{z-\hat{H}}|i;N\rangle\\
&=\frac{1}{z-{E_{i,fs}}-N\hbar\omega-\langle
i;N|\Sigma^*_{i;N}(z)|i;N\rangle},
\end{split}
\label{greenfunction}
\end{eqnarray}
with the proper self-energy
\begin{eqnarray}
\begin{split}
\Sigma^*_{i;N}(z)&\approx\Sigma^{(0)}_{i;N}(z)+\Sigma^{(2)}_{i;N}(z),
\end{split}
\label{greenfunction}
\end{eqnarray}
where $\Sigma^{(0)}_{i;N}(z)$ and $\Sigma^{(2)}_{i;N}(z)$ are the
parts containing the 0th and 2nd order of $\hat{H}_e$, respectively.
Here we have neglected higher orders of atom-photon interactions,
due to the reasons presented in the discussion.

\begin{figure}[h]
\includegraphics [height=3cm,width=8cm]{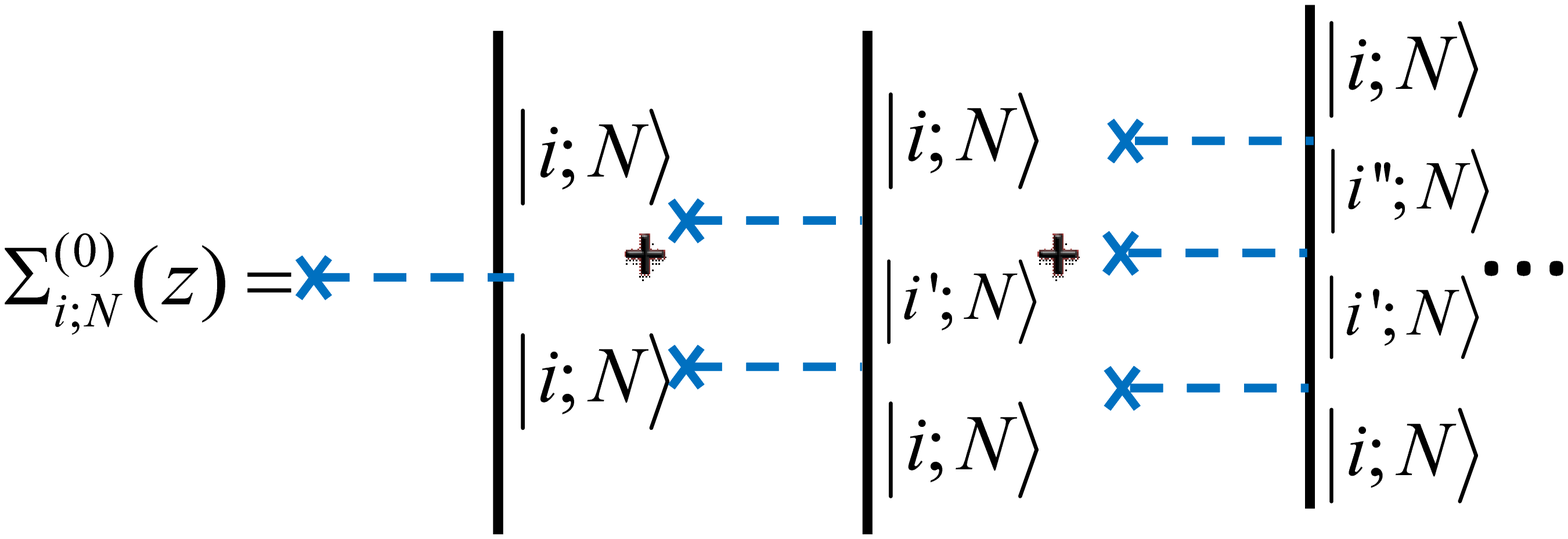}
\caption{ (Color online)Diagrammatic representation of the proper
self-energy $ \Sigma^{(0)}_{i;N}(z)$. $|i'\rangle$ and $|i''\rangle$
denote different eigenstates of $\hat{H_{fs}}$ which have the same
parity with $|i\rangle$ but with different $J$ or principle quantum
numbers. Every solid line(black) marked by $|i;N\rangle$ represents
a factor $G^{0}_{i;N}(z)=1/(z-\langle i;N|\hat{H_0}|i;N\rangle)$,
and every dashed line(blue) marked by $|i;N\rangle$ and
$|i';N\rangle$ denote a factor $\langle
i;N|\hat{H}_{hfs}|i';N\rangle=\langle i|\hat{H}_{hfs}|i'\rangle$. A
summation is performed over the index $i',i''...$ in the end.}
\label{rep_diagram12}
\end{figure}

\begin{figure}[h]
\includegraphics [height=6cm,width=8cm]{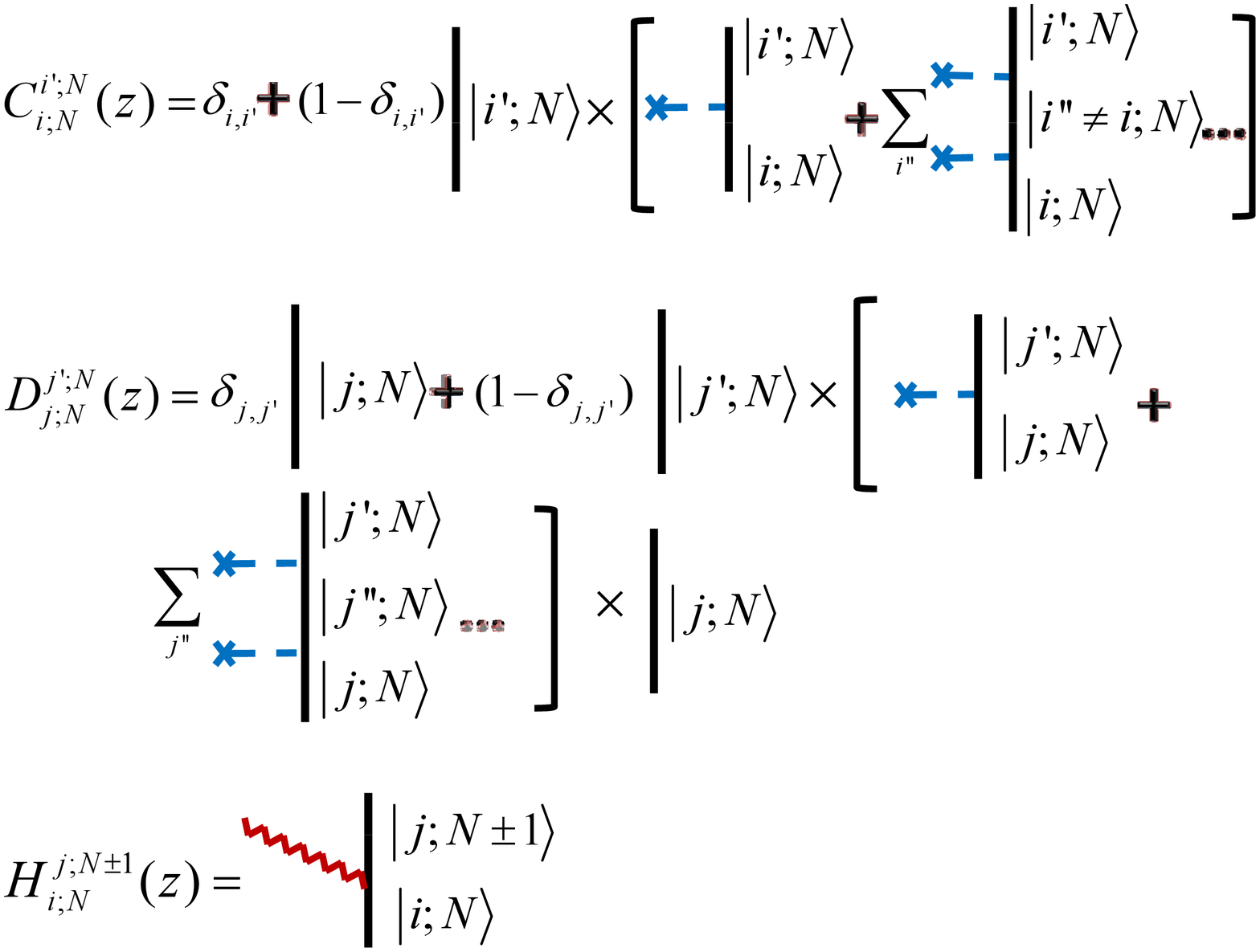}
\caption{ (Color online)Diagrammatic representation of the $C$, $D$
and $H$ factors. All the $|i\rangle,|j\rangle$ denote eigenstates of
$\hat{H_{fs}}$. Every solid line(black) marked by $|i;N\rangle$
represents a factor $G^{0}_{i;N}(z)=1/(z-\langle
i;N|\hat{H_0}|i;N\rangle)$ and every dashed line(blue) marked by
$|i;N\rangle$ and $|i';N\rangle$ denotes a factor $\langle
i'|\hat{H}_{hfs}|i\rangle$. In $H$ factor, the wiggly line(red)
marked by $|i;N\rangle$ and $|j;N\pm 1\rangle$ represents a factor
$\langle j;N\pm 1|\hat{H}_e|i;N\rangle$. Here $N$ is the photon
number of the radiation field, and $|j\rangle,|j'\rangle...$
represent states having a parity opposite to that of $|i\rangle$.}
\label{rep_diagram2}
\end{figure}

The terms in $\Sigma^{(0)}_{i;N}(z)$ and $\Sigma^{(2)}_{i;N}(z)$ can
also be represented diagrammatically. In Fig. (\ref{rep_diagram12}),
$\Sigma^{(0)}_{i;N}(z)$ has a similar structure to that of
$\sigma^*_{i,hfs}(z)$ except the solid lines represent
$G^{0}_{i;N}(z)=1/(z-\langle i;N|\hat{H_0}|i;N\rangle)$, not
$g^{0}_{i}(z)=1/(z-\langle i|\hat{H_{fs}}|i\rangle)$. The diagrams
of $\Sigma^{(2)}_{i;N}(z)$ are more complicated. It can be
decomposed into three kinds of factors: $C$, $D$ and $H$, which are
presented in Fig. (\ref{rep_diagram2}):
\begin{eqnarray}
\begin{split}
\Sigma _{i;N}^{(2)}(z) &= \sum_{\substack{{i}',{i}''\\{j}',{j}''}}
{C_{i;N}^{{i}'';N}\cdot H_{{j}'';N \pm 1}^{{i}'';N}\cdot D_{{j}';N
\pm 1}^{{j}'';N \pm 1}\cdot H_{{i}';N}^{{j}';N \pm 1}\cdot
C_{{i};N}^{{i}';N}}
\end{split}
\label{decompose}
\end{eqnarray}

As shown in the diagram, the hyperfine interactions are included in
the $C$ and $D$ factors, meanwhile the $H$ factor contains the
atom-photon interactions. The factors $C$ and $D$ are very similar,
except in $C^{i';N}_{i;N}$, $|i\rangle$ cannot be an intermediate
state, meanwhile in $D_{i;N}^{i';N}$ there is no such restriction.
Using the notation $/\kern-0.47em \delta_{ij}=1-\delta_{ij}$ for
convenience, we can calculate the $C$, $D$ and $H$ factors formally:
\begin{eqnarray}
\begin{split}
C_{i;N}^{i';N}&=\delta_{i,i'}+/\kern-0.47em \delta_{i,i'}\frac{1-G^{0}_{i;N}\Sigma^{(0)}_{i;N}}{G^{0}_{i;N}}D_{i;N}^{i';N}\\
D_{j;N}^{j';N}&=[\delta_{j,j'}+/\kern-0.47em
\delta_{j,j'}C_{j;N}^{j';N}]
\frac{G^{0}_{j;N}}{1-G^{0}_{j;N}\Sigma^{(0)}_{j;N}}\\
H_{i;N}^{j;N \pm 1}&=\langle j;N\pm 1|\hat{H}_e|i;N\rangle
\end{split}
\label{factors}
\end{eqnarray}

\subsection{The poles of the Green's function}
Now we calculate the poles of the Green's function Eq.
(\ref{greenfunction}). Using Eq. (\ref{thezrelation}), we can see
that
\begin{eqnarray}
\begin{split}
E_{i;hfs}=\langle i;N|\Sigma^{(0)}_{i;N}(z)|i;N\rangle,
\end{split}
\label{E0}
\end{eqnarray}
therefore $E_{i;N}^{(0)}=E_{i;fs}+N\hbar\omega+E_{i;hfs}$ is the
eigenenergy given the zeroth order approximation in atom-photon
interaction. Because the variations of $\Sigma^{(2)}_{i;N}(z)$ with
$z$ are much slower than those of $G_{i;N}(z)$, in particular near
$z=E_{i;N}^{(0)}$, we can use the relation
\begin{eqnarray}
\begin{split}
\Sigma^{(2)}_{i;N}(z)\approx \Sigma^{(2)}_{i;N}(E_{i;N}^{(0)}).
\end{split}
\end{eqnarray}
The factors $C_{i;N}^{i';N}(z)\approx C_{i;N}^{i';N}(E_{i;N}^{(0)})$
and $D_{j;N}^{j';N}(z)\approx D_{j;N}^{j';N}(E_{i;N}^{(0)})$ can be
calculated perturbatively using the series in Fig.
(\ref{rep_diagram2}):
\begin{eqnarray}
\begin{split}
&C_{i;N}^{i';N}(E_{i;N}^{(0)})=\delta_{i,i'}+/\kern-0.47em
\delta_{i,i'}\mathcal{O}(\frac{\langle
i|\hat{H}_{hfs}|i'\rangle}{\Delta E_{i,i'}}),\\
&D_{j;N\pm1}^{j';N\pm1}(E_{i;N}^{(0)})
=\frac{1}{\Delta E_{i,j}\pm\hbar\omega}[\delta_{j,j'}+/\kern-0.47em
\delta_{j,j'}\mathcal{O}(\frac{\langle
j|\hat{H}_{hfs}|j'\rangle}{\Delta E_{i,j'}\pm\hbar\omega})],
\end{split}
\label{factorCD1}
\end{eqnarray}
where $\mathcal{O}$ gives the approximation order, and $\Delta
E_{i,j}=E_i-E_j$ denote the energy difference between hyperfine
states $|i\rangle$ and $|j\rangle$. At the orders of approximation
$OC=\mathcal{O}(\langle i|\hat{H}_{hfs}|i'\rangle/\Delta E_{i,i'})$
for the C factor and $OD=\mathcal{O}(\langle
j|\hat{H}_{hfs}|j'\rangle/(\Delta E_{i,j'}\pm\hbar\omega))$ for the
D factor, we have
\begin{eqnarray}
\begin{split}
C_{i;N}^{i';N}(E_{i;N}^{(0)})&\approx\delta_{i,i'},\\
D_{j;N\pm1}^{j';N\pm1}(E_{i;N}^{(0)})&\approx\delta_{j,j'}\frac{1}{E_{i}-E_{j}\pm\hbar\omega}.
\end{split}
\label{factorCD2}
\end{eqnarray}

\subsection{The formula of $E^{[2+\epsilon]}$ method }
Taking Eq. ({\ref{factorCD2}}) into Eq. (\ref{decompose}) and after
further simplification:
\begin{eqnarray}
\begin{split}
&\Sigma^{(2)}_{i;N}(E_{i;N}^{(0)})=\sum_{j}{\frac{|\langle
i;N|\hat{H}_{e}|j;N+1\rangle|^2}{\Delta
E_{i,j}-\hbar\omega}+(\omega\rightarrow -\omega)}\\&=-{3 \pi c^2
I_{L}}\sum\limits_{j\neq
i}\frac{A_{J_{ji}}(2F_j+1)(2F_i+1)(2J_j+1)\omega_{F_{ji}}}{\omega^3_{J_{ji}}(\omega^2_{F_{ji}}
-\omega^2)}\\&\times \left(\begin{array}{ccc} F_{j}& 1 &
F_i\\M_{F_j}
&p&-m_{F_i}\end{array}\right)^2\left\{\begin{array}{ccc} J_i& J_j&
1\\F_j&F_i &I\end{array}\right\}^2,
\end{split}
\label{lightshift1}
\end{eqnarray}
where $I_L=(\hbar N \omega)/(\epsilon_0V)$ is the light intensity,
$\hbar\omega_{J_{ji}}=E_{j,fs}-E_{i,fs}$,
$\hbar\omega_{F_{ji}}=E_{j}-E_{i}$ and
\begin{eqnarray}
A_{J_{ji}} =
\frac{e^2}{4\pi\epsilon_0}\frac{{4\omega_{J_{ji}}^3}}{{3\hbar
{c^3}}}\frac{1}{{2{J_j} + 1}}|\left\langle{{j}} \left\| \hat{d}
\right\| {{i}}\right\rangle|^2. \label{EinsteinCo}
\end{eqnarray}
is the Einstein coefficient for the fine structure transition
between $|i\rangle$ and $|j\rangle$. The term inside curly brackets
is a $6J$ symbol, meanwhile the large round brackets is the $3J$
symbol which describes the selection rules and relative strength of
the transitions. Here $p$ stands for the polarization of the
trapping laser ($p=1,0,-1$ for right-hand, linear and left-hand
polarization, respectively).

As shown in Eq. (\ref{lightshift1}), we are able to calculate the ac
Stark shift as long as we have the data for the Einstein
coefficients $A_{J_{ji}}$, the energies $\omega_{F_{ji}}$ and
$\omega_{J_{ji}}$. For many elements, there are theoretical and
experimental data available for these physical quantities.
Furthermore, the differential light shift can be obtained by
comparing the absolute light shift of two hyperfine levels of
interest. In Sec.III, we apply formula (14) to each of the ground
state hyperfine doublet of Cs or Rb and subtract the two to obtain
the differential light shift. A comparison with the numerical result
of TOPT is also presented.


\section{Calculation for Microwave Clock Transition Levels}
We calculate the light shifts with $E^{[2+\epsilon]}$ method and TOPT
for two alkali elements, Cs and Rb, both of which have been proposed
as potential choices for microwave optical lattice
clocks~\cite{rev}. In alkali microwave atomic clocks, the clock
transition is between the lower hyperfine level
$|nIJF_1M_{F_1}\rangle$ and the upper hyperfine level
$|nIJF_2M_{F_2}\rangle$ of the ground state with magnetic sublevels
$M_{F_1} = M_{F_2} = 0$. The transition frequency shift for the
microwave clock is
\begin{eqnarray}
\delta \nu_{clock}=(E^{[2+\epsilon]}_{F_2M_{F_2}} -
E^{[2+\epsilon]}_{F_1M_{F_1}})/h.
\end{eqnarray}
\subsection{Numeric results}
\begin{figure}[h]
\begin{center}
\includegraphics [width=8cm]{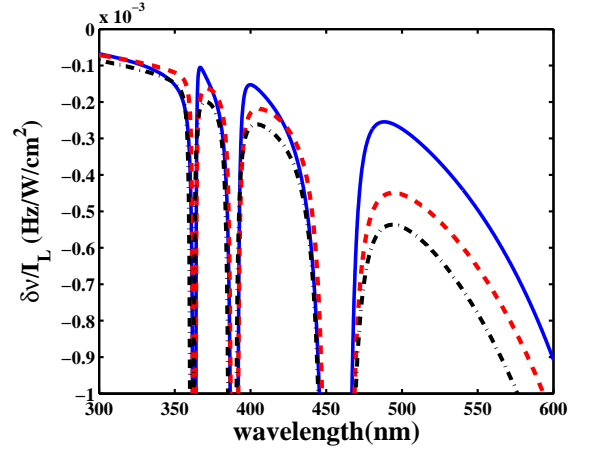}
\end{center}
\caption{(Color online) Wavelength dependence of the differential ac
Stark shift for the ground state hyperfine doublet of Cesium133.
TOPT result using GraspVU data (blue solid), $E^{[2+\epsilon]}$ method
result using GraspVU data (red dashed) and $E^{[2+\epsilon]}$ method
result using experiment data (black dot-dashed) are presented. The
trapping laser wavelength is ranging from 300nm to 600nm.}
\label{Cs1}
\end{figure}

\begin{figure}[h]
\begin{center}
\includegraphics [width=8cm]{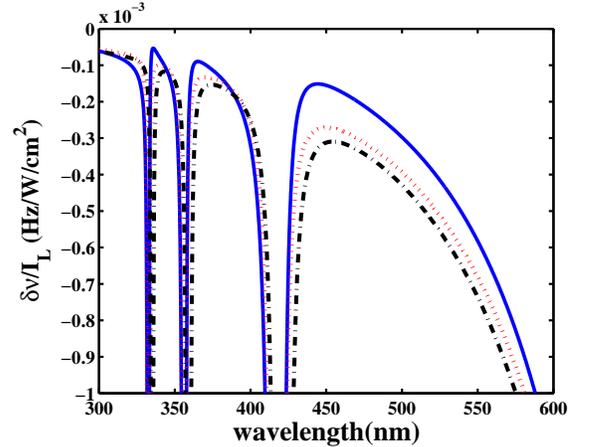}
\end{center}
\caption{(Color online) Wavelength dependence of the differential ac
Stark shift for the ground state hyperfine doublet of Rubidium87.
TOPT result using GraspVU data (blue solid), $E^{[2+\epsilon]}$
method result using GraspVU data (red dashed) and $E^{[2+\epsilon]}$
method result using experiment data (black dot-dashed) are
presented. The trapping laser wavelength is ranging from 300nm to
600nm.} \label{Rb1}
\end{figure}

First, as elaborated in Appendix A, we use the GraspVU
program~\cite{Parpia} to construct the database of atomic structures
of Rb and Cs, including the wavefunctions, energy levels, hyperfine
interactions and electric dipole transition strength. With this
database, we can calculate the differential ac Stark shift with
$E^{[2+\epsilon]}$ method and TOPT. The computation results at various
wavelength are presented in Fig.(\ref{Cs1}) and Fig.(\ref{Rb1}). We
can see both the TOPT values(blue solid line) and $E^{[2+\epsilon]}$
method values(red dashed line) stay negative for all the wavelength
and they are very close to each other.

One advantage of $E^{[2+\epsilon]}$ method is that instead of
constructing atomic structure data, we can use the existing
experimental values of Einstein coefficients and hyperfine energies
in Eq.(\ref{lightshift1}). As an illustration, we use experimental
values for the hyperfine splittings~\cite{HFSSplit} and transition
rates ~\cite{NIST,CsEin}. The calculation results are shown in
Fig.(\ref{Cs1}) and Fig.(\ref{Rb1}) in black dot-dashed lines. As we
can see, they also stay negative for all the wavelength.


Second, we can compare the differential ac Stark shift at specific
trapping light wavelength with experimental values. Using TOPT and
GraspVU database, the differential shifts at 780 nm and 532 nm with
linear polarized light are $-2.00\times 10^{-2} Hz/mW/cm^2$ and
$-3.99\times 10^{-4} Hz/mW/cm^2$, while using $E^{[2+\epsilon]}$
method and GraspVU database, the differential shifts are
$-1.92\times 10^{-2} Hz/mW/cm^2$ and $-5.59\times 10^{-4}
Hz/mW/cm^2$. Both of those results are in agreement with the
experimental values $-2.27\times 10^{-2} Hz/mW/cm^2$ and
$-3.51\times 10^{-4} Hz/mW/cm^2$~\cite{P.Rosenbusch}. The
$E^{[2+\epsilon]}$ method with experiment database gives $-2.15\times
10^{-2} Hz/mW/cm^2$ and $-6.68\times 10^{-4} Hz/mW/cm^2$, also in
agreement with experiments.


\subsection{Order of approximation}
In Eq. (\ref{factorCD1}), we have shown that our result is an
approximation at the order $OC$ for C factors and $OD$ for D
factors. For light shift of ground states of Cesium and Rubidium,
$OC$ is the ratio of $\hat{H}_{hfs}$ matrix elements between the
ground state and a higher $S_{1/2}$ state to their energy
difference, meanwhile $OD$ equals the ratio of $\hat{H}_{hfs}$
matrix elements between two $P$ states to the light detuning. These
error terms are very small in typical experiment conditions. Take Rb
for example. When $|i\rangle$ is the hyperfine doublet of the ground
state, $\Delta E_{i,i'}$ is at least $6*10^{14}Hz$ meanwhile
${\langle i|\hat{H}_{hfs}|i'\rangle}$ is at most $2*10^9Hz$, which
makes $OC\thicksim \mathcal{O}(10^{-5})$. On the other hand, the
largest $\langle j|\hat{H}_{hfs}|j'\rangle$ is between $5p$ and $6p$
which is less than $3*10^8Hz$, therefore for a wide range of
trapping laser wavelength, $OD\thicksim \mathcal{O}(10^{-5})$ can be
satisfied. For Cs, a similar estimation can be made. The C factor is
approximated at the order
$\mathcal{O}(3*10^{9}Hz/5*10^{14}Hz)\thicksim \mathcal{O}(10^{-5})$,
and for D factor, the largest $\langle j|\hat{H}_{hfs}|j'\rangle$ is
between $6p$ and $7p$ which is about $4*10^8Hz$, therefore for a
wide range of wavelength of the light field, our method is also
valid at the order of accurate $\mathcal{O}(10^{-5})$.

However, when we calculate the differential light shift between two
hyperfine levels of the ground state, we need to reevaluate the
order of accuracy since the light shifts of these two levels are
very close to each other. After further investigation, we discover
that for our method to be valid, it requires:
\begin{eqnarray}
\begin{split}
&\max_j |\frac{\Delta E_{i,hfs}{d_{ij}}}{\Delta
E_{i,j}-\hbar\omega}|\gg\max_{j,j'}|\frac{\langle
j|\hat{H}_{hfs}|j'\rangle{d_{ij'}}}{\Delta E_{i,j'}-\hbar\omega}|,\\
&\max_j |\frac{\Delta E_{i,hfs}d_{ij}^2}{\Delta
E_{i,j}-\hbar\omega}|\gg\max_{j',i'}|\frac{\langle
i|\hat{H}_{hfs}|i'\rangle d_{ij'}d_{i'j'}}{\Delta E_{i,i'}}|,
\end{split}
\label{breakdownlimit}
\end{eqnarray}
where $|i\rangle$ is the hyperfine level of the ground state,
$\Delta E_{i,hfs}$ is the corresponding hyperfine splitting, and
$d_{ij}=\langle i\|\hat{d}\|j\rangle$ is the reduced matrix element
of electric dipole operator.

Again take Rb for example. In a far-off-resonant laser field, the
first inequality can hold since $\Delta E_{i,hfs}\approx 7*10^{9}Hz$
meanwhile the largest $|\langle j|\hat{H}_{hfs}|j'\rangle|<
4*10^8Hz$. In the second inequality, because the largest
$|d_{ij}^2/d_{ij'}d_{i'j'}|\ge 5$ and as stated above, $|{\langle
i|\hat{H}_{hfs}|i'\rangle}/\Delta E_{i,i'}|\thicksim
\mathcal{O}(2*10^9Hz/6*10^{14}Hz)$, it requires the detuning
$|\Delta E_{i,j}-\hbar\omega|\ll 10^{16}Hz$. A similar analysis can
be performed for Cs. This explains why our calculation results of
differential light shift are close to the TOPT results, and when the
detuning gets bigger, the difference between two results also
increases. In conclusion, for a wide range of wavelength of the
trapping laser, our method is also valid for calculating the
differential light shift of the ground state hyperfine doublet.




\section{Discussions and conclusions}
In Sec.II, we have derived the $E^{[2+\epsilon]}$ method using the
Green's function. In order to obtain the simple expression in
Eq.(\ref{lightshift1}) for the ac Stark shift, we have performed a
partial summation of the original perturbation expansion of the
proper self-energy $\Sigma^*_{i;N}(z)$. The largest contribution to
the light shift is the second order process
$\Sigma^{*(2)}_{i;N}(z)$. The lowest order diagram
$\Sigma_{omit}(z)$ that has been omitted is the fourth order
atom-photon interaction, which has an order of magnitude:
\begin{eqnarray}
\begin{split}
&\Sigma_{omit}(z)/(\Sigma^{*(2)}_{i;N}(z))^2\approx1/\Delta,\\
\end{split}
\label{ordercomparison}
\end{eqnarray}
where $\Delta$ is the detuning of the radiation field. Take Rb for
example. For a wavelength $\lambda=850nm$ laser field with $20E_R$
trap depth, $\Delta\sim10^{13}Hz$ and
$\Sigma^{*(2)}_{i;N}(z)\sim10^5Hz$, so the ratio
$\Sigma_{omit}(z)/\Sigma^{*(2)}_{i;N}(z)$ is much smaller than
$10^{-5}$ and we can
keep the atom-photon interaction at the second order. 



The calculation results shown in Fig.(\ref{Cs1}) and Fig.(\ref{Rb1})
suggest that there is no magic wavelength for Rb and Cs microwave
clocks. However, at certain wavelength of trapping laser, the shift
difference is very small and varies rather slowly versus the
wavelength, implying the differential light shifts would be very
stable against the trapping light's intensity and frequency
fluctuations. These wavelength are thus suitable for atom trapping
in precision measurement. As listed in Table.I, the recommended
wavelength obtained using both methods and different databases of
atomic structure are very close to each other.

\begin{table}[ht]
\begin{flushleft}
\caption{Using TOPT with graspVU data, $E^{[2+\epsilon]}$ with graspVU
data and $E^{[2+\epsilon]}$ with experiment data (Expt.), the
recommended wavelength $\lambda$ of trapping laser for Cs and Rb
microwave lattice clocks are calculated and listed in the table.}
\end{flushleft}
\centering
\begin{tabular}{|c|c|c|c|c|c|}\hline\hline
\multicolumn{3}{|c|}{Cs}&\multicolumn{3}{||c|}{Rb}\\\hline
\multicolumn{1}{|c|}{TOPT}&\multicolumn{1}{c|}{$E^{[2+\epsilon]}$}&\multicolumn{1}{c|}{$E^{[2+\epsilon]}$}&\multicolumn{1}{||c|}{TOPT}&\multicolumn{1}{c|}{$E^{[2+\epsilon]}$}&\multicolumn{1}{c|}{$E^{[2+\epsilon]}$}\\
\multicolumn{1}{|c|}{(GraspVU)}&\multicolumn{1}{c|}{(GraspVU)}&\multicolumn{1}{c|}{(Expt.)}&\multicolumn{1}{||c|}{(GraspVU)}&\multicolumn{1}{c|}{(GraspVU)}&\multicolumn{1}{c|}{(Expt.)}\\\hline
366.8nm&371.2nm&370.4nm&\multicolumn{1}{||c|}{355.8nm}&340.0nm&343.0nm\\\hline
399.8nm&405.2nm&404.0nm&\multicolumn{1}{||c|}{365.2nm}&370.2nm&374.0nm\\\hline
488.2nm&494.2nm&493.8nm&\multicolumn{1}{||c|}{444.4nm}&450.0nm&454.2nm\\\hline
871.2nm&870.2nm&874.8nm&\multicolumn{1}{||c|}{785.4nm}&784.8nm&788.2nm\\\hline\hline
\end{tabular}
\label{I}
\end{table}

In summary, we have used Green's function to derive the
$E^{[2+\epsilon]}$ method Eq. (\ref{lightshift1}) for calculating
the ac Stark shift of hyperfine levels. A discussion about the
approximation orders of Eq. (\ref{factorCD1}) is made and we
discover that for a wide range of trapping laser wavelength,
$E^{[2+\epsilon]}$ method is valid both for calculating the absolute
and the differential light shift, which is also indicated by the
numerical results. Moreover, the experimental values for atomic
levels and Einstein coefficients can be utilized in
$E^{[2+\epsilon]}$ method. This implies that $E^{[2+\epsilon]}$
method can be applied to other group elements for
which \textit{ab initial} atomic structure databases are difficult to construct.\\

\section*{ACKNOWLEDGMENTS}
We thank Thibault Vogt for having carefully reviewed our article.
This work is partially supported by the state Key Development
Program for Basic Research of China (No.2011CB921501) and NSFC (No.11334001), RFDP (No.20120001110091).\\

\appendix

\section{Numerical calculation of atom structure data}

We use the GraspVU code~\cite{Parpia} which is based on the
relativistic multi-configuration Dirac-Fock (MCDF)
method~\cite{MCDF} to compute the energy levels and generate the
wave functions. In the following, we first present a brief summary
of the MCDF method and our calculation strategies. For an
$N$-electron atom or ion, the Dirac-Coulomb Hamiltonian can be
expressed as (in atomic unit)
\begin{eqnarray}
\hat{H}_{DC} = \sum\limits_{i = 1}^N {[c\alpha {p_i} + (\beta -
1){c^2} - \frac{Z}{{{r_i}}}] + \sum\limits_{i = 1}^{N - 1}
{\sum\limits_{j = i + 1}^N {|{r_i} - {r_j}{|^{ - 1}}} } }.
\end{eqnarray}
So the eigenvalue problem is
\begin{eqnarray}
 \hat {H}_{DC}\left| {\Gamma PJM}
\right\rangle  = {E_\Gamma }\left| {\Gamma PJM} \right\rangle.
\end{eqnarray}
Where $\left| {\Gamma PJM} \right\rangle$ represents the atomic
state functions (ASFs) with $\Gamma$ denoting other quantum numbers.
The ASF can be written as linear combinations of configuration state
functions (CSFs) with the same parity $P$, total angular momentum
$J$ and magnetic quantum number $M$ ,

\begin{eqnarray}
\left| {\Gamma PJM} \right\rangle  = \sum\limits_{i = 1}^{{n_c}}
{{C_{i} ^{\Gamma }}\left| {{\gamma_i }PJM} \right\rangle }.
\end{eqnarray}

Here ${C_i}^\Gamma$  is the mixing coefficients, $\gamma_i$
represents all information required to define CSF uniquely, and
$n_{c}$ is the number of CSFs. The CSFs $\left| {\gamma_i PJM}
\right\rangle$ which form a quasi-complete basis set in Hilbert
space are linear combinations of Slater determinants of order $N$
constructed from atomic orbital functions (AOs). By applying
variational method to Eq.(B2), we can obtain the mixing coefficients
and the AOs self-consistently.
\begin{table}[ht]
\begin{flushleft}
\caption{Calculation strategies of the atomic orbital sets for Cs
and Rb.}
\end{flushleft}
\centering
\begin{tabular}{|c|c|c|c|c|c|c|c|}\hline\hline
&step 1 \textsuperscript{\emph{a}}&step 2&step 3&step 4&step 5&step
6 \textsuperscript{\emph{b}}&step 7\\\hline
Cs&1*,2*,3*,4s&6s,6p&7s,7p&8s,4f&9s,5f&$\widetilde{10s}$,$\widetilde{10p}$
&$\widetilde{11s}$,$\widetilde{11p}$\\
&4p,4d,5s,5p&5d      &  6d    &  8p,7d      & 9p,8d       &     $\widetilde{9d}$,$\widetilde{6f}$     &    $\widetilde{10d}$,$\widetilde{7f}$       \\
\hline
Rb&1*,2*,3*&5s,5p&6s,6p&7s,7p&8s,4f&$\widetilde{9s}$,$\widetilde{9p}$
&$\widetilde{10s}$,$\widetilde{10p}$\\
&4s,4p & 4d     &  5d    &   6d     & 8p,7d       &     $\widetilde{8d}$,$\widetilde{5f}$     &   $\widetilde{9d}$,$\widetilde{6f}$        \\
\hline\hline
\end{tabular}
\begin{flushleft}
\textsuperscript{\emph{a}} $1*$ represents 1$s$ spectroscopy
orbital, $2*$ represents 2$s$,2$p$ spectroscopy orbitals while
$3*$ represents 3$s$,3$p$ and 3$d$ spectroscopy orbitals\\
\textsuperscript{\emph{b}} $\widetilde{nl}$ represents pseudo
orbital
\end{flushleft}
\end{table}
Our calculations are based on our recently proposed
multi-configuration self-consistent field (MCSCF)
strategies~\cite{bqing,shchen}. We use GraspVU code ~\cite{Parpia}
to optimize a set of high-quality orbital basis where pseudo
orbitals~\cite{Gao,bqing,shchen,cheng,zhang} are included, by which
we can take into account the electron correlation effects
adequately. The pseudo orbitals determined by variational method are
specific linear combinations of infinite bound type orbitals and
continuum orbitals. More specifically, the occupied orbitals are
obtained from ground state single configuration calculations. The
other AOs are obtained through MCSCF calculations with all the core
orbitals fixed. For Cs, we first extend the AOs classified by
manifold $\nu$ of effective quantum numbers to 9$s$, 9$p$, 8$d$,
5$f$ by optimizing the orbitals with same angular momentum $l$
separately. These orbitals are treated as spectroscopy orbitals
(labeled as $nl$ and with fixed number of radial nodes, i.e.
$n-l-1$), and they represent the physical state. Configurations are
generated by single excitation from $6s^1$ and all the energy levels
are optimized. Note that these spectroscopy orbitals are adequate
for polarizability calculations in present frequency scale. In order
to consider electron correlations adequately, we further extend the
AOs as pseudo orbitals (labeled as $\widetilde{nl}$  and without
restriction on radial nodes) to $\widetilde{11s},\widetilde{11p},
\widetilde{10d},\widetilde{7f}$ by optimizing the orbitals with same
angular momentum $l$ separately. With all the spectroscopy orbitals
included, the configurations are generated by single and double
excitations only allowing one electron excited from 5$p$ or 6$s$
orbitals. The optimized energy levels are the same as in the
spectroscopy orbital calculations. We call the basis set which
satisfies the desired accuracy of calculations quasi-complete basis
set, as listed in Table.II. The quasi-complete basis set for Rb is
constructed in the same way as Cs.

All the calculations in this work depend on the quasi-complete basis
obtained in the previous procedure. To verify the quality of this
basis, we first perform configuration interaction calculations (CI)
with configuration generated by single and double excitations from
$5p^{6}6s^{1}$. With Breit interactions included, the fine-structure
energy levels of $ns$, $np$($5\leq n \leq 9$ for Cs, $4\leq n \leq
8$ for $Rb$) states agree with the experimental results within
$1\%$, both for Cs and Rb. Hence, the correlations are considered
adequately and the ASFs should be adequate for further calculation.
By using these ASFs, the transition parameters and transition matrix
elements for electric dipole transitions are calculated, and the
hyperfine interaction such as the magnetic dipole interaction (A
constants) and electric quadrupole interaction (B constants) are
computed. The off-diagonal hyperfine interaction elements on the
$|F,M_{F}\rangle$ basis in Eq. (\ref{hyperfinestate}) can also be
obtained from RHFS ~\cite{rhfs}.


\bibliographystyle{elsarticle-num}

\end{document}